A library-based Monte Carlo technique enables rapid equilibrium sampling of a protein model with atomistic components.

Artem B. Mamonov, Divesh Bhatt, Derek J. Cashman, Daniel M. Zuckerman

Department of Computational Biology, School of Medicine, University of Pittsburgh, Pittsburgh, Pennsylvania

Corresponding author: D. M. Zuckerman, Tel.: 412 648-3335; E-mail: ddmmzz@pitt.edu

**Abstract**

There is significant interest in rapid protein simulations because of the time-scale limitations of all-atom methods. Exploiting the low cost and great availability of computer memory, we report a Monte Carlo technique for incorporating fully flexible atomistic protein components (e.g., peptide planes) into protein models without compromising sampling speed or statistical rigor. Building on existing approximate methods (e.g., Rosetta), the technique uses pre-generated statistical libraries of all-atom components which are swapped with the corresponding protein components during a simulation. The simple model we study consists of the three all-atom backbone residues – Ala, Gly, and Pro – with structure-based (Gō-like) interactions. For the five different proteins considered in this study, LBMC can generate at least 30 statistically independent configurations in about a month of single CPU time. Minimal additional cost is required to add residue-specific interactions.

## I. Introduction

Proteins are tiny machines that perform their function by motion in many cases. It has been recently recognized that proteins evolved not only to have a specific structure but also dynamic properties to perform certain biological functions. Conformational fluctuations are believed to be important for many protein functions such as signaling (1), catalysis (2,3), and intracellular transport (4). Indeed, experimental techniques now allow observation of protein motions at a wide range of different time scales (5-7). However, the simultaneous determination of structure and dynamics remains a challenging task (8) and has motivated the development of theoretical methods.

Among theoretical methods for studying protein dynamics molecular dynamics (MD) simulation is one of the most popular methods, due to its all-atom realism and the reasonable accuracy of current forcefields. However, a considerable gap remains between the timescale of many biologically important motions and the timescale accessible to MD simulations – see, e.g., Ref. (9) about the attempted folding of a WW domain. Numerous efforts have been directed toward developing new atomistic algorithms to bridge this timescale gap (10-14).

Coarse-grained (CG) and multi-resolution approaches for modeling proteins are gaining in popularity (15-19). Recently our group developed the "resolution exchange" (ResEx) algorithm that utilizes a ladder of several different coarse-grained models with simulations simultaneously running at different levels and occasional exchange attempts between different levels (20,21). In the more efficient top-down implementation (20) ResEx relies on converged sampling at the most coarse-grained level and gradual "de-coarsening" of generated configurations to higher resolution levels. Beyond equilibrium sampling, it can be anticipated that CG models will be needed for path sampling of dynamic processes (22).

The need for converged coarse-grained sampling led us to test how simplified a protein model should be in order to be fully sampled in reasonable CPU time. We developed a coarse-grained model based on rigid peptide-plane geometry incorporated into a united-residue model (23). Our group also developed a robust convergence analysis that directly probes the fundamental configuration space distribution (24). Convergence analysis of trajectories generated with the rigid-planes model indicated that about 50 statistically independent configurations can be generated in less than a week of single CPU time for a protein containing about 90 residues. However, the model can be applied only to proteins containing less that 100 residues due to the difficulty of finding near-native rigid-plane starting configurations (23).

The present study improves our previous model by incorporating all-atom backbone flexibility with united-residue interactions at a small computational cost and without compromising statistical rigor. This is achieved by using an apparently novel library-based Monte Carlo (LBMC) technique which exploits pre-generated libraries of peptide-plane configurations, although the formalism applies to arbitrary fragment libraries. The technique is successfully applied to several proteins, the largest of which contains more than 300 residues. The convergence analysis indicates that full sampling is easily obtained in less than a month of single processor CPU time.

## II. Library-based Monte Carlo (LBMC)

The library-based Monte Carlo approach is very general and not limited to the present model consisting of peptide-plane fragments. Libraries consisting of full residues, or multi-residue fragments, could readily be employed. Although the well-known folding program Rosetta (25) uses *ad hoc* libraries (not corresponding to a known distribution), in principle adaptations could be made to such a program to generate canonical, Boltzmann-factor sampling.

Library-based techniques should be of increasing interest in the future, with the fast growth and declining cost of computer memory. Intuitively, it seems clear that vast computer resources are devoted to redundant calculations (e.g. of bond energy) which could more parsimoniously be retrieved from memory.

Library-based Monte Carlo attempts to minimize calculations by splitting a molecule into fragments with a library of configurations for each fragment generated in advance. Such a strategy is natural for proteins, which consists of only 20 building blocks. During a simulation, fragment configurations in the present state of the system are swapped with configurations in the library and the new state is accepted according to the corresponding acceptance criterion derived below. Computational efficiency can be achieved by pre-calculating the interactions within a fragment which do not have to be calculated again during a simulation. The libraries should be generated only once and then can be used in repeated simulations. The idea of using libraries of pre-calculated configurations has been around for some time in the polymer growth field and several library-based algorithms have previously been suggested (26-28).

Our technique is partly inspired by the success of the fragment assembly software Rosetta, in which short fragments of known protein structures are assembled into native-like conformation using *ad hoc* library-based Monte Carlo (25). Rosetta is one of the most successful methods for *de novo* protein structure prediction (29). The main difference between our technique compared to Rosetta is that LBMC is statistically rigorous and allows the generation of statistical ensembles and calculation of thermodynamic properties. In principle our technique can be also used to aid protein structure prediction since the native state will be sampled in an ensemble based on a forcefield of sufficient quality.

As in all Monte Carlo (MC) simulations, different types of trial moves can be used alternately or in combination. For example, one can make hybrid trial moves in which fragment configurations are swapped along with making random changes to the degrees of freedom which are not included with the libraries. If desired, all degrees of freedom can be pre-sampled in libraries.

The acceptance criterion for a library-based swap move depends on the distribution of the library. If the degrees of freedom in the fragment are already distributed according to the Boltzmann factor (for those coordinates alone), the acceptance criterion becomes particularly simple and independent of fragment energy as shown below. However, as we also discuss, in some cases it may be useful to employ other library distributions.

**II.A Defining fragments and the energy decomposition**

To be fully general, we first consider an arbitrary division of a molecule into $M$ non-overlapping fragments. On the individual fragment level, a configuration of fragment $i$ containing $N_i$ degrees of freedom is described by $\vec{r}_i = \{x_{i1},...,x_{iN_i}\}$. Such fragments need not correspond to groups of atoms. For instance, in our case, we have excluded the $\psi$ backbone dihedral angle from our peptide-plane fragments – and all of the $\psi$ dihedrals together can be considered to constitute a single fragment. Regardless of the division into fragments, the usual total potential energy of the system (e.g., from a standard forcefield) can then be decomposed as

$$U^{\text{tot}}(\vec{r}_1,...,\vec{r}_M) = \sum_{i=1}^{M} U_i^{\text{frag}}(\vec{r}_i) + U^{\text{rest}}(\vec{r}_1,...,\vec{r}_M), \qquad (1)$$

where $U_i^{\text{frag}}(\vec{r}_i)$ contains all energy terms which depend solely on the coordinates of fragment $i$ but with no interactions between fragments. $U^{\text{rest}}(\vec{r}_1,...,\vec{r}_M)$ simply includes the balance of the energy terms, and thus encompasses all interactions between fragments and any energy terms dependent on the coupling between fragments. In a typical molecular system, $U_i^{\text{frag}}(\vec{r}_i)$ will contain all van der Waals, electrostatics and bonded terms *internal* to the fragment.

**II.B Monte Carlo simulation with libraries**

The acceptance criteria necessary for library-based MC can be derived in more than one way. In an appendix we present derivations based on a continuum description, while the present section uses statistical ideas more directly relevant to the discrete nature of the libraries.

Our present derivation is based on the fact that LBMC is nothing more than the use of Metropolis MC to perform re-weighting – or more precisely "resampling" (30). This can be understood based on standard simulation ideas.

First, why do we need to re-weight? Quite simply, the configuration space spanned by our libraries has a built-in bias. The library for fragment $i$ consists of a set of configurations distributed according to some function $p_i^{\text{frag}}(\vec{r}_i)$, which will be specified below. If we draw a random configuration from each library, we will generate full molecular configurations distributed according the product of the single-fragment distributions,

$$p^{\text{lib}}(\vec{r}_1,...,\vec{r}_M) \propto \prod_{i=1}^{M} p_i^{\text{frag}}(\vec{r}_i). \qquad (2)$$

In any realistic system, this simple-product distribution will differ from the desired equilibrium distribution $p^{\text{eq}}$, which in our case is proportional to the usual Boltzmann factor of the full potential, Eq. 1, namely,

$$p^{\text{eq}}(\vec{r}_1,...,\vec{r}_M) \propto \exp\left[-\beta U^{\text{tot}}(\vec{r}_1,...,\vec{r}_M)\right]. \qquad (3)$$

At a minimum, the full Boltzmann factor will contain terms which couple the fragments and which are absent from the fragment libraries by construction. Regardless of the precise differences, the simple-product distribution of Eq. 2 will need to be reweighted in order to recover the correct distribution.

The ideas of reweighting and resampling are textbook subjects (30) and can readily be understood. The basic idea is that one has an incorrect distribution, $p^{lib}(\mathbf{r})$, instead of the desired target distribution $p^{eq}(\mathbf{r})$, with $\mathbf{r} = (\vec{r}_1,...,\vec{r}_M)$ in our case. This means that the calculation of a canonical average in the $p^{eq}$ ensemble requires that every configuration from the $p^{lib}$ distribution be assigned a weight (31)

$$w(\mathbf{r}) = p^{eq}(\mathbf{r}) / p^{lib}(\mathbf{r}). \tag{4}$$

The denominator exactly cancels the incorrect frequency from the $p^{lib}$ distribution. We note that overall normalization is irrelevant in a weighted average. Another way to understand the weights of Eq. 4, is by observing that a partition function can always be re-written using a "sampling function" *g*. Specifically, the identity:

$$Z = \int [d\mathbf{r}] e^{-\beta U_{tot}(\mathbf{r})} = \int [g(\mathbf{r}) d\mathbf{r}] e^{-\beta U_{tot}(\mathbf{r})} / g(\mathbf{r}), \tag{5}$$

where the bracketed function denotes either uniform or *g*-distributed sampling, implies the weights of configurations must be $e^{-\beta U_{tot}(\mathbf{r})}$ or $e^{-\beta U_{tot}(\mathbf{r})}/g(\mathbf{r})$, respectively. In our case $g = p^{lib}$.

"Resampling" describes the procedure for generating a representative ensemble of *unweighted* configurations in the $p^{eq}$ ensemble from a sample of $p^{lib}$ (30). Quite simply, configurations from an ensemble distributed according to $p^{lib}$ should be included in the $p^{eq}$ ensemble with probability proportional to the weight $w(\mathbf{r})$. Operationally, resampling can be performed in a number of different ways. As an example, the simplest method examines each configuration **r** in the original $p^{lib}$ sample and accepts or rejects it with probability $w(\mathbf{r})/w_{max}$, where $w_{max}$ is the maximum weight among all sampled configurations.

Direct resampling as just described is effectively impossible in our case, because of the astronomical number of full molecular configurations under consideration. In a typical case, we will employ over $10^5$ configurations *per fragment library,* with fragments representing individual amino acids. Thus, for a protein of 100 residues, there are over $10^{500}$ possible configurations, and we could never consider every one.

Metropolis Monte Carlo is designed precisely for such a situation, where all possible configurations cannot be considered directly. Further, due to the finite precision of digital computing, ordinary "continuum" MC is, in fact, discrete. Thus, standard MC can be seen as a method to sample a set of Boltzmann-factor-distributed configurations based on a much larger "library" of finely and uniformly discretized Cartesian configurations (32). One can call this "sampling" or even "resampling," but the net effect is to transform one distribution into another. Our goal is the same. The key difference in our case is that we wish to sample our discrete space of *biased* library-based configurations with probability proportional to *w* of Eq. 4. The word "biased" here really only means that our initial distribution is neither uniform nor equilibrium.

Based on this logic, it is straightforward to adapt Metropolis MC to the necessary resampling. Ordinary MC samples a distribution *p* based on the condition of detailed balance, namely,

$$p(o)p_{gen}(o \to n)p_{acc}(o \to n) = p(n)p_{gen}(n \to o)p_{acc}(n \to o), \qquad (6)$$

where $p_{gen}$ and $p_{acc}$ are the usual generating and acceptance conditional probabilities, while "$o$" and "$n$" are shorthand for the old and new (trial) configurations, respectively. The usual case of setting $p = p^{eq}$ corresponds to (re) sampling the "library" of finite-precision configurations into the Boltzmann distribution, as already noted.

In our case, we want to set $p = w$ in Eq. 6, in order to re-weight our library distribution, which has been customized for molecular fragments. In the resampling case, then, the correct acceptance criterion becomes:

$$p_{acc}(o \to n) = \min\left[1, \frac{w(n)p_{gen}(n \to o)}{w(o)p_{gen}(o \to n)}\right], \qquad (7)$$

where overall normalization of $w$ is irrelevant as in standard Metropolis MC. Our Results section carefully demonstrates the validity of this acceptance criterion in a simple, verifiable system.

It is important to note that, as in standard Metropolis Monte Carlo, once the distribution or weighting function is selected for Eqs. 6 and 7, arbitrary trial moves among the library configurations can be employed, so long as any asymmetries in generating probabilities are accounted for. Further, whenever the generating probability is symmetric, i.e., when $p_{gen}(i \to j) = p_{gen}(j \to i)$ for all $i$ and $j$, we obtain the simpler form

$$p_{acc}(o \to n) = \min\left[1, \frac{w(n)}{w(o)}\right]. \qquad (8)$$

The acceptance criterion of Eq. 8 is related to "Metropolized independence sampling," originally derived by Hastings (32).

**II.C Library Monte Carlo with molecular fragments**

In order to employ library-based Monte Carlo with our molecular fragment libraries, we need to derive the appropriate weight function of Eq. 4 for use in the acceptance criterion of Eq. 8. As just described, this weight function is the ratio of the targeted equilibrium distribution of Eq. 3 to the distribution based on the pre-calculated fragment configurations. The fragment-based distribution of Eq. 2 results from separate libraries for each fragment in the present case. In our study, library configurations $\vec{r}_i$ for each fragment $i$ will be distributed according to the "internal" Boltzmann factors, implying:

$$p_i^{frag}(\vec{r}_i) \propto \exp\left[-\beta U_i^{frag}(\vec{r}_i)\right], \qquad (9)$$

where the fragment energy has been defined in Eq. 1. In essence, the "global" library of full configurations we are considering is constructed from independent configurations for each fragment, and is therefore distributed according to the simple product of fragment distributions of Eq. 9.

We can now directly calculate the necessary weight function 4 based on Eqs. 2, 3 and 9, which yields:

$$w(\vec{r}_1, ..., \vec{r}_M) = \exp\left[-\beta U^{rest}(\vec{r}_1, ..., \vec{r}_M)\right]. \qquad (10)$$

This is the weight function which should be employed in the acceptance criteria of Eqs. 7 and 8 when using libraries distributed as in Eq. 9. Thus, the acceptance probability for trial moves with symmetric generating probability reduces to the simple expression

$$p_{acc}(o \to n) = \min\left[1, \exp\left(-\beta \Delta U^{rest}\right)\right], \quad (11)$$

where $\Delta U^{rest} = U^{rest}(n) - U^{rest}(o)$. The appendix provides a more traditional derivation of this acceptance criterion based on detailed-balance arguments, which is equivalent to "Metropolized independence sampling" (30) originally derived by Hastings (32).

By design, all of the fragment energies have cancelled out from Eq. 11, which can serve two purposes. First, the cancellation significantly reduces the number of interactions which need to be calculated in testing for acceptance via Eq. 8. Second, because all interactions internal to the fragments have been pre-sampled, one can hope that the discretized configuration space embodied in the libraries has more overlap with the target distribution than would a uniform sampling implicit in the uniform trial moves of a typical Monte Carlo simulation. That is, trial moves based on library swaps will automatically account for interactions internal to the swapped fragment.

For completeness, we note that the $\psi$ dihedral angles, which constitute one of our fragments, are treated as standard continuum variables in the present study. This is a minor technical point, and is equivalent to placing the associated energy terms in $U^{rest}$ rather than as a separate fragment term. No bias results from this choice, and indeed it underscores the freedom to employ libraries for some variables but not others.

**II.D Neighbor lists in library-based Monte Carlo.**

The MC scheme described above has poor control of the acceptance ratio for trial moves where fragment configurations are selected at random from one of the fragment libraries. We found that with peptide-plane fragment libraries the acceptance ratio is very small for a reasonable size protein (100 residues or more). The reason is the "small angle disaster" – i.e., even a difference of a few degrees in angle accumulated over one fragment may result into large conformational change 20-30 Å away from the swapped fragment. To improve the acceptance ratio, library configurations can be classified into neighbor lists so that each configuration can be assigned a number of similar neighbor configurations. Instead of selecting configurations randomly from the library they can be selected from the neighbor list, thereby increasing the acceptance ratio.

We thus need to determine the two generating probabilities in Eq. 7 which correspond to using neighbor lists. For simplicity, we are solely interested in trial moves which change *a single fragment configuration* – i.e., the overall configurations "o" and "n" differ only in a single fragment. That is, we wish to consider "neighbors" of a given configuration within a single library. (Generalizations to multi-fragment trial moves are straightforward, but not presented here.)

Fortunately, once a list of neighbors has been determined for a given fragment configuration (as described below), the generating probabilities take a trivial form. Indeed, $p_{gen}(i \to j)$ answers the question, "What is the probability the *computer program* will choose configuration *j* as a trial move, given that *i* is the present configuration?" If configuration $i$ has $k_i$ neighbors, of which one is selected at random, then:

$$p_{\text{gen}}(i \to j) = 1/k_i. \tag{12}$$

Regardless of how the library was generated, once the neighbor list is determined, the (conditional) generating probability takes this simple form.

The acceptance criterion appropriate for neighbor lists is therefore given by

$$p_{\text{acc}}(o \to n) = \min\left[1, \frac{w(n)/k_n}{w(o)/k_o}\right] = \min\left[1, \exp(-\beta \Delta U^{\text{rest}})\frac{k_o}{k_n}\right], \tag{13}$$

where $k_o$ must be understood as the number of neighbors for the particular (single) fragment configuration selected for a trial move, and $k_n$ is the number of neighbors of the trial fragment configuration. The criterion in Eq. 13 is also derived from a continuous picture in the Appendix.

If neighbor lists are constructed to have the same number of neighbors for all configurations in a given library, then the acceptance criterion in Eq. 13 reduces to the simpler symmetric form of Eq. 11. In the present study, all neighbor lists were constructed to be the same size, as we now describe.

In practice, fragment configurations can be classified into neighbor lists using some metric describing the similarity of configurations to each other. This metric can be, for example, RMSD or the sum of absolute differences over all backbone bond and dihedral angles. This metric can be calculated between all pairs of configurations in the library and sorted based on similarity to each other. When building the neighbor list, configurations most similar to the chosen one should be selected to be in its neighbor list. Further, to satisfy microscopic reversibility, if configuration $i$ has in its neighbor list configuration $j$ then $j$ must have $i$. Of course, the case when the same configuration occurs multiple times in the same neighbor list should be excluded. One problem with the neighbor lists is that they may form isolated clusters. This problem can be solved by alternating the local neighbor list swap moves with the global moves when library configurations are selected randomly from the whole library.

**II.E Using libraries with non-Boltzmann distributions.**

We found that, for some systems, libraries distributed according to the Boltzmann factor are not effective because fragment configurations may have very low population densities in high-energy transition regions of configuration space that may be important in a full system. For example, high free energy regions connecting the left ($\varphi<0$) and right ($\varphi>0$) sides of the Ramachandran plot are very rarely populated in equilibrium libraries of amino acids. However, these regions are functionally important for some proteins and ignoring them would make it difficult to sample potentially important configurations. Inefficiency also arises when a fragment library has more configurations than necessary in the low free energy regions.

With these points in mind, the fragment libraries can be improved by over-representing some regions of configuration space and under-representing others. In principle, the fragment configurations can be biased toward known protein structures similar to the Rosetta method (25). When biased, configurations must be assigned weights based on the ratio of the desired library size to the true (unbiased) equilibrium populations. If every library configuration $j$ is classified into some state $s(j)$ – i.e., some region of configuration space – then its weight is

$$b(j) = \frac{P_{bias}(s(j))}{P_{eq}(s(j))}, \qquad (14)$$

where $P_{bias}(s)$ is the biased library size (fractional population) for state $s$ and $P_{eq}(s)$ is the true equilibrium population in the same state. Note that $P_{eq}(s)$ is simply proportional to the unbiased library counts for the fragment $i$ under consideration, as well as to the local partition function $Z(s) = \int_{\vec{r}_i \in s} d\vec{r}_i \exp\left[-\beta U_i^{frag}(\vec{r}_i)\right]$. We assume that within each state, a library is unbiased – i.e. configurations are distributed according to the Boltzmann factor. The states can represent, for example, different regions of the Ramachandran plot. When the biased library is used, the library distribution of Eq. 2 should be modified to account for the introduced bias

$$p^{lib}(\vec{r}_1,...,\vec{r}_M) \propto \prod_{i=1}^{M} b(\vec{r}_i) \exp\left[-\beta U^{frag}(\vec{r}_i)\right]. \qquad (15)$$

This library distribution can be used along with the target distribution of Eq. 3 to find the necessary weighting function:

$$w(\vec{r}_1,...,\vec{r}_M) = \frac{\exp\left[-\beta U^{rest}(\vec{r}_1,...,\vec{r}_M)\right]}{\prod_{i=1}^{M} b(\vec{r}_i)}. \qquad (16)$$

Using this weighting function, the acceptance criterion for a swap move with biased libraries can be obtained

$$p_{acc}(o \rightarrow n) = \min\left[1, \exp\left(-\beta \Delta U^{rest}\right) \prod_{i=1}^{M} \frac{b(\vec{r}_i^o)}{b(\vec{r}_i^n)}\right]. \qquad (17)$$

Note that the product of biasing weights needs to be calculated only for fragments which were swapped in the current MC move. For the rest of the fragments (which were not swapped) the biasing weights are the same and cancel out.

**II.F Practical library generation and the use of dummy atoms**

Libraries of fragment configurations can be generated using any standard canonical sampling method, for example, Langevin dynamics or Metropolis MC. The only statistical requirement for a library is that it should represent the true equilibrium distribution – that is, it must be consistent with Eq. 9 – for all the degrees of freedom in the fragment.

*Dummy atoms*. In practice, the efficiency of LBMC depends on how the system is split into fragments and what degrees of freedom are included in each fragment. Because fragments are sampled separately and independently from each other it is usually convenient to include and sample the extra six "dummy" degrees of freedom that specify the orientation of fragments relative to each other. However, because the dummy atoms are not considered part of the fragment, their interactions with the fragment atoms should not contribute to $U_i^{\text{frag}}(\vec{r}_i)$. This can be achieved by making the dummy atoms non-interacting i.e., by setting their van der Waals parameters and partial charges to zero.

As noted, our peptide plane fragments include all of the relative degrees of freedom (with respect to the subsequent fragment) except for the $\psi$ dihedral. The $\psi$ dihedral was not included with the planes to allow high density of configurations needed for generation of neighbor lists. We found that when $\psi$ dihedral was included with peptide planes, it was very hard to find "good" neighbor configurations because the density of configurations significantly decreased in the expanded (by $\varphi$-$\psi$ plane) configuration space. In practice to allow incorporation of a separate Ramachandran potential in the future, the peptide plane libraries were modified to be conditional on the $\varphi$ dihedral angle – i.e., uniform in $\varphi$ with a suitable energy correction. In future studies, we plan to use fragments based on residues (rather than peptide planes) which will permit a more natural treatment of the $\varphi$ and $\psi$ dihedrals.

As will be discussed below three peptide plane types were employed in our protein model corresponding to Ala, Gly and Pro residues. In our simulations the library size used was $2.9*10^5$ for Ala, $2*10^5$ for trans-Pro, $2*10^5$ for cis-Pro and $3.6*10^5$ for Gly. The cis-Pro library was implemented to allow the sampling of proteins containing Pro residue in cis conformation. All peptide planes corresponding to cis-Pro were swapped only with the cis-Pro library and trans-Pro planes were swapped only with the trans-Pro library. For all our libraries the neighbor lists were generated to contain 10 configurations.

### III. Protein Model

In our previous work we showed that complete sampling of configuration space is possible with a simple rigid peptide plane protein model in several weeks of single CPU time (23). Building on our previous work we further improve the model by including all-atom based backbone flexibility at a small computational cost.

In our protein backbone model, a fragment is represented by a peptide plane configuration containing all of the atomic backbone degrees of freedom except $\psi$ which is sampled as a standard continuous variable (see above). The planes are coupled through interactions sited, for simplicity, at alpha carbons, as described below. Our protein backbone model is schematically shown in Figure 1.

Three different types of peptide planes are used, corresponding to Ala, Gly and Pro residues needed for the correct atomic model of the backbone. All non-Gly and non-Pro residues are reduced to "pseudo-Ala" plane containing a beta carbon without hydrogen atoms, which is a natural choice for a backbone model since all non-Gly and non-Pro residues have similar Ramachandran map propensities (33). For the Pro residues all of the ring atoms are included with the peptide plane because they affect backbone configurations. Libraries of peptide plane configurations were generated using Langevin dynamics as implemented in the Tinker v. 4.2 software package (http://dasher.wustl.edu/tinker/)(34) with OPLS-AA forcefield (35) and implicit GB/SA solvent (36) at 298 K. The libraries and corresponding energies are available on our website: www.epdb.pitt.edu.

To stabilize the native state and also allow fluctuations, this initial study employs "structure-based" or Gō interactions among alpha carbons (37,38). Our previous studies showed that the Gō potential can reproduce reasonable protein fluctuations compared to experimental data (23,39). Neighboring peptide planes are excluded from Gō interactions so that residue *i* can interact only starting from residue *i+3*. The $C_\alpha$ interaction centers for the Gō potential consist of native and non-native interactions. Thus the total potential energy not internal to the fragments (peptide-planes) is

$$U^{rest} = U^{G\bar{o}}_{nat} + U^{G\bar{o}}_{non}, \qquad (18)$$

where $U^{G\bar{o}}_{nat}$ is the total energy for native contacts and $U^{G\bar{o}}_{non}$ is the total energy for non-native contacts. All residues that are separated by a distance less than a cutoff, $R_{cut}$, in the experimental structure are given native interaction energies defined by a square well

$$U^{G\bar{o}}_{nat} = \sum_{i<j}^{native} u^{nat}(r_{ij})$$

$$u^{nat}(r_{ij}) = \begin{cases} \infty & \text{if } r_{ij} < r^{nat}_{ij}(1-\delta) \\ -\varepsilon & \text{if } r^{nat}_{ij}(1-\delta) \leq r_{ij} < r^{nat}_{ij}(1+\delta) \\ 0 & \text{otherwise} \end{cases}, \qquad (19)$$

where $r_{ij}$ is the distance between $C_\alpha$ atoms of residues *i* and *j*, $r^{nat}_{ij}$ is the distance between residues in the experimental structure, $\varepsilon$ determines the energy scale of Gō interactions, and $\delta$ sets the width of the square well. All residues that are separated by more than $R_{cut}$ in the experimental structure are given non-native interaction energies defined by

$$U^{G\bar{o}}_{non} = \sum_{i<j}^{non-native} u^{non}(r_{ij})$$

$$u^{non}(r_{ij}) = \begin{cases} \infty & \text{if } r_{ij} < (\rho_i + \rho_j)(1-\delta) \\ h\varepsilon & \text{if } (\rho_i + \rho_j)(1-\delta) \leq r_{ij} < R_{cut} \\ 0 & \text{otherwise} \end{cases}, \qquad (20)$$

where $\rho_i$ is the hard-core radius of residue *i*, defined at half the $C_\alpha$ distance to the nearest non-covalently bonded residue in the experimental structure, and *h* determines the strength of the repulsive interactions.

For this study, parameters were chosen to be similar to those in Ref. (23), i.e., $\varepsilon=1.0$, $h=0.3$, $\delta=0.2$, and $R_{cut}=8.0$ Å.

## IV. Results and Discussion

*Toy system*. We performed a "reality check" of the LBMC technique by applying it to a simple one-dimensional toy system represented by two particles connected to harmonic springs. Two cases were considered: in the first case, the particles are non-interacting, whereas in the second case particles interact via a repulsive Coulombic ($r^{-1}$) potential. For each case, the probability density function (pdf) along the one-dimensional coordinate for each particle was calculated using LBMC and checked using standard Brownian dynamics (BD). For LBMC a library of $10^6$ configurations was generated in advance for a system composed of one particle connected to a harmonic spring. During LBMC simulation both particles were sampled using the same library. For BD simulations a standard over-damped Langevin Dynamics procedure without velocities was used (40). The results for LBMC and BD are compared in Figure 2B for non-interacting particles and in Figure 2C for interacting particles. The agreement between LBMC and BD for both cases is excellent. Note that in Figure 2C the equilibrium distance between particles increased due to Coulomb repulsion that was correctly reproduced by LBMC.

In preliminary tests on poly-alanine systems using the full OPLS-AA forcefield, we obtained excellent agreement with equilibrium distributions from Langevin simulations. These results will be described elsewhere.

*Test proteins.* To compare our LBMC technique with our previously developed rigid peptide plane model (23) we performed LBMC calculations for the same three proteins previously studied with rigid peptide planes. These were the binding domain of protein G (PDB code 1PGB, residues 1-56), the N-terminal domain of calmodulin (PDB code 1CLL, residues 4-75), and barstar (PDB code 1A19, residues 1-89).

As in our previous study we chose the dimensionless simulation temperature $k_BT/\varepsilon$ to be slightly below the unfolding temperature of the protein. The unfolding temperatures for each protein were determined via short simulations of $5*10^7$ MC steps for 13 different temperatures. The temperatures determined for the production runs turned out to be the same as used in the previous study i.e. $k_BT/\varepsilon = 0.5$ for protein G, $k_BT/\varepsilon = 0.4$ for calmodulin and $k_BT/\varepsilon = 0.6$ for barstar.

Each protein system was first equilibrated for $10^8$ MC steps followed by the production runs of $2*10^9$ MC steps. For the production runs frames were saved at the interval of 1000 MC steps, generating equilibrium ensembles of $2*10^6$ frames. The LBMC simulations were performed on a single Xeon 3.6 GHz CPU and it took 33 hours to complete the production run for protein G, 48 hours for calmodulin, and 52 hours for barstar.

The root mean square deviations (RMSD) along the trajectory relative to the experimental structure for three different proteins are shown in Figure 3. All backbone heavy atoms were used for RMSD calculations. These plots demonstrate the ability of our model to sample large conformational fluctuations of proteins along with the apparent convergence of the trajectories characterized by the stable behavior, in contrast to typical MD plots.

A quantitative convergence analysis was performed using the method reported in Ref. (24), i.e. convergence was analyzed by studying the variance of the structural-histogram bin populations. The analysis determines the necessary time between trajectory frames so that statistically decorrelated behavior occurs – i.e. "the decorrelation time". Figure 4 shows the convergence properties of LBMC simulations based on the ratio $\sigma^2$ of the average population variance to that expected for independent sampling. A normalized variance ($\sigma^2$) of one indicates statistical independence. The resulting decorrelation times are reported in Table 1 and indicate that the LBMC technique allows high quality sampling (ca. 100 decorrelation times) in several weeks of single CPU wall-clock time.

*CDC25B*. As discussed above, one of the main advantages of the flexible peptide-plane model compared to a rigid plane model is that it can be used for proteins containing more that 100 residues. Thus we applied LBMC to human CDC25B catalytic domain (PDB code 1QB0, residues 374-550) containing 177 residues. This is a dual-specificity phosphatase with established links to cancer (41).

Simulation parameters were similar to those for test proteins. The simulation temperature, like for other proteins, was chosen slightly below the unfolding temperature based on 5 different temperature simulations and was determined to be $k_BT/\varepsilon$=0.5. The system was equilibrated for $2*10^9$ MC steps followed by the production run of $5*10^9$ MC steps with frames saved every $10^4$ frames generating the equilibrium ensemble of $5*10^5$ frames. It took ca. 13 days of a single Xeon 3.6 GHz CPU wall-clock time to complete the production run.

The trajectory of RMSD values relative to the experimental structure for CDC25B is shown in Figure 3D. The RMSD based on the whole protein (black line in Figure 3D) shows very large conformational fluctuations. Inspection of trajectory configurations reveals that the C-terminus helix (residues 525-550) is flexible and partially unfolds during the simulation. The RMSD calculated based on the stable part of the protein (residues 374-524) is shown as a red line in Figure 3D and indicates that the rest of the protein is stable. Twenty different configurations along the trajectory are superposed in Figure 5 and show that the C-terminus helix is unstable (in our model) and partially unfolds during the simulation along with the stable behavior of the rest of the protein. The convergence analysis of the LBMC trajectory is shown in Figure 4D and the corresponding structural decorrelation time is reported in Table 1, indicating that converged sampling is possible in several weeks of a single CPU wallclock time.

It is worth noting that the crystal structures of CDC25 isoforms A and B are very similar except for the C-terminus region (corresponding to residue 529 and beyond in CDC25B) that is unfolded in isoform A (42). In isoform B this region forms an alpha-helix lying along the protein body with an anion binding site at the end occupied by one $Cl^-$ to stabilize the electrostatic charge in this region. Since the protein molecules in the crystal are in contact by C-termini, the stability of this region may be a crystal structure artifact caused by crystal packing forces and favorable electrostatic interactions (43).

In practice for some applications our statistical backbone ensembles can be converted into *ad hoc* all-atom ensembles. We explored this possibility by converting 500 configurations of CDC25B generated with LBMC to all-atom configurations by adding side chains using the program SCWRL 3.0 (44). To remove steric clashes the side chain addition was followed by energy minimization using the OPLS-AA forcefield (35). These structures are available at www.epdb.pitt.edu. Inspection of the generated configurations revealed that ca. 40% of configurations have the disulfide bond formed between residues Cys426 and Cys473 which is not present in the original crystal structure of CDC25B (PDB code 1QB0). A literature search revealed that that these Cys residues are conserved in that phosphatase family and easily oxidize to form a disulfide bond (42,43). It was speculated that formation of this disulfide bond may be important for self-inhibition during oxidative stress. The fact that our model can sample backbone conformations optimal for formation of the disulfide bond observed experimentally can be considered as a partial validation of our model.

*GGBP*. To test the limits of our LBMC technique we applied it to an even larger protein, the D-Galactose/D-glucose binding protein (GGBP) (PDB code 2GBP, residues 1-309) containing 309 residues. This is one of the first proteins used experimentally to study equilibrium fluctuations directly (45), and the details of these fluctuations will be examined in a future study.

As with other proteins, the simulation temperature was chosen slightly below the unfolding temperature based on 13 short simulations of $3*10^8$ MC steps. The simulation temperature was determined to be $k_BT/\varepsilon =0.7$. For the production run the system was run for $3*10^9$ MC steps with frames saved every $10^4$ MC steps, generating an equilibrium ensemble of $3*10^5$ frames. It took ca. 22 days of single Xeon 3.6 GHz CPU wallclock time to complete the production run.

The RMSD plot relative to the experimental structure is shown in Figure 3E. This plot demonstrates the ability of our technique to sample large conformational fluctuations for a large protein. The convergence analysis of the LBMC trajectory is shown in Figure 4E and the decorrelation time is reported in Table 1, indicating that a converged sampling is possible in about a month of single CPU time. A detailed analysis of the experimentally studied conformational fluctuations will be reported elsewhere.

## V. Conclusions

We developed and tested a novel library-based Monte Carlo (LBMC) technique that allows the efficient incorporation of all-atom flexibility of protein fragments at a small computational cost compared to much simpler models, and without compromising statistical rigor. We applied LBMC to extend our previous "united residue" model consisting of rigid peptide-planes (23) so that full atomic backbone flexibility was included.

We successfully modeled proteins containing more than 300 residues, greatly exceeding the practical limit of 100 residues found with our previously developed rigid peptide-plane model (23). The results demonstrate the capability of our new model and technique to sample reasonable conformational fluctuations. Importantly a quantitative analysis demonstrates that fully sampled equilibrium ensembles can be obtained in about a month of single CPU time.

The main difference between the present LBMC technique compared to other fragment based methods (e.g. Rosetta) is that LBMC is statistically rigorous and allows the generation of statistical ensembles and calculation of thermodynamic properties.

Although all-atom based flexibility is present in our simple test model, it lacks side-chains and hence accurate physico-chemical interactions. Because our model can achieve converged sampling in such a short time, it readily can be enhanced by more accurate interactions. Additional potential energy terms such as Ramachandran propensities, hydrogen-bonding and hydrophobic interactions are currently being studied.

Potential applications of backbone ensembles generated with LBMC include docking and homology modeling. To make accurate predictions, these methods critically rely on ensembles of configurations, especially when substrate binding induces large conformational changes. For this purpose fully atomistic MD and MC simulations have been used before (46,47) but have practical limitations due to their computational cost and time scale limitations. Since our model can rapidly sample large conformational fluctuations and transitions, it may be useful for generating ensembles of configurations suitable for methods like docking and homology modeling. One group is currently investigating the suitability of our CDC25B ensembles for docking (Arantes G. M., personal communications 2008).

Further technical improvements certainly are possible. For instance, additional computation time may be saved by storage of interactions among neighboring residues. Larger fragments should facilitate more localized (crankshaft-like) trial moves. Also, the use of a fine grid (39) can permit storage of interactions among non-bonded residues.

**Appendix**

The acceptance criteria for trial moves which are library swaps can also be derived by incorporating the library distribution into $p_{gen}$. Let us start from the standard equation of detailed balance

$$p^{eq}(o) p_{gen}(o \to n) p_{acc}(o \to n) = p^{eq}(n) p_{gen}(n \to o) p_{acc}(n \to o), \qquad (21)$$

where the trial configuration is fully described by the set of new fragment configurations, $n = \{\vec{r}_1^{\,n}, ..., \vec{r}_M^{\,n}\}$, and the old configuration is similarly given by $o = \{\vec{r}_1^{\,o}, ..., \vec{r}_M^{\,o}\}$. As in Eq. 3, $p^{eq}$ is the Boltzmann factor of the full potential, while $p_{gen}$ and $p_{acc}$ are the conditional probabilities for generating and accepting trial moves, as above. From Eq. 21 the usual general expression for the acceptance criterion can be obtained, namely

$$p_{acc}(o \to n) = \min\left[1, \frac{p^{eq}(n) p_{gen}(n \to o)}{p^{eq}(o) p_{gen}(o \to n)}\right]. \qquad (22)$$

*Swap with a full fragment library.* Let us first consider the case when a library-based trial move is the replacement ("swap") of a single fragment configuration with a random choice the *full* library of the same fragment $i$. That is, we are choosing the trial configuration $n = \{\vec{r}_1^{\,o}, \vec{r}_2^{\,o}, ..., \vec{r}_{i-1}^{\,o}, \vec{r}_i^{\,n}, \vec{r}_{i+1}^{\,o}, ..., \vec{r}_M^{\,o}\}$, which differs from the configuration $o$ by only a single fragment $\vec{r}_i^{\,n}$. But further, because the trial fragment configuration is chosen *independently* of the old configuration, the generating probability will depend only on the probability of choosing the new fragment from the library. In our case, each library is distributed according to the fragment Boltzmann factor, Eq. 9, so that the generating probability for a swap with the full library of fragment $i$ is therefore

$$p_{gen}(o \to n) = \frac{\exp\left[-\beta U_i^{frag}(\vec{r}_i^{\,n})\right]}{Z_i}, \qquad (23)$$

which is indeed independent of $o$. Here, $Z_i = \int_{V_i} d\vec{r}_i \exp\left[-\beta U_i^{frag}(\vec{r}_i)\right]$ is the normalizing partition function for fragment $i$, with limits of integration over the full hypervolume $V_i$ of configuration space available to the fragment. We note that the strategy of choosing a trial configuration independent of the old configuration was originally suggested by Hastings (32), and is called "Metropolized independence sampling" (30).

The ratio of generating probabilities is obtained by a similar analysis of the reverse move, leading to the result

$$\frac{p_{gen}(n \to o)}{p_{gen}(o \to n)} = \frac{\exp\left[-\beta U_i^{frag}(\vec{r}_i^{\,o})\right]}{\exp\left[-\beta U_i^{frag}(\vec{r}_i^{\,n})\right]}, \qquad (24)$$

where the normalizing partition function has cancelled because the same fragment library is considered in both cases.

The full acceptance criterion can be derived by employing the ratio of Eq. 24 along with the equilibrium distribution of Eq. 3. Recalling the decomposition of the total potential energy of Eq. 1, *all fragment energy terms cancel in the final acceptance criterion.* The fragment terms aside from that of the swapped fragment cancel trivially because their configurations are unchanged, but the ratio of Eq. 24 leads – by design – to the cancellation of even the fragment term *i* with the equilibrium Boltzmann factor. The net result is that

$$p_{acc}(o \to n) = \min\left[1, \exp(-\beta \Delta U^{rest})\right], \tag{25}$$

for a swap generated using Eq. 23, where $\Delta U^{rest} = U^{rest}(n) - U^{rest}(o)$. Eq. 25 is the same as Eq. 11 derived using the re-weighting approach.

*Swap based on a neighbor list.* Now let us consider a more general case, namely, when a trial fragment configuration is selected from a pre-generated neighbor list instead from the whole library. In a continuum description, this amounts to selecting a configuration $\vec{r}_i^n$ from a pre-defined region of configuration space $V_i^o$ "neighboring" the old fragment configuration $\vec{r}_i^o$. (Note that the choice of the region $V_i^o$ is arbitrary, and it could be disconnected.) In our case, a trial fragment configuration will be selected according to the fragment Boltzmann factor solely within $V_i^o$, which corresponds to the generating probability

$$p_{gen}(o \to n) = \frac{\exp\left[-\beta U_i^{frag}(\vec{r}_i^n)\right]}{Z_{i,o}} = \frac{\exp\left[-\beta U_i^{frag}(\vec{r}_i^n)\right]}{\int_{V_i^o} d\vec{r}_i \exp\left[-\beta U_i^{frag}(\vec{r}_i)\right]}, \tag{26}$$

which can be contrasted with Eq. 23.

To construct the ratio of generating probabilities, we must recognize that the normalizing local partition functions $Z_{i,o}$ and $Z_{i,n}$ (for $p_{gen}(o \to n)$ and $p_{gen}(n \to o)$, respectively) will not be the same now because generally the neighborhoods will differ, i.e., $V_i^o \neq V_i^n$. The ratio of generating probabilities thus becomes, instead of Eq. 24,

$$\frac{p_{gen}(n \to o)}{p_{gen}(o \to n)} = \frac{\exp\left[-\beta U_i^{frag}(\vec{r}_i^o)\right]/Z_{i,n}}{\exp\left[-\beta U_i^{frag}(\vec{r}_i^n)\right]/Z_{i,o}} \tag{27}$$

We note that to satisfy microscopic reversibility, the neighborhoods should be defined so that if $V_i^o$ contains $\vec{r}_i^n$, then $V_i^n$ should contain $\vec{r}_i^o$.

The full acceptance criterion for neighbor lists can be derived by substituting Eqs. 3 and 27 into Eq. 22. Initially, one finds

$$p_{acc}(o \to n) = \min\left[1, \exp(-\beta \Delta U^{rest}) \frac{Z_{i,o}}{Z_{i,n}}\right], \tag{28}$$

but we want to eliminate the partition functions and convert the criterion for the discrete library case of interest. This can be done using the key fact that, because each fragment library is Boltzmann-distributed over its *entire configuration space* as in Eq. 9, the number of library configurations $k_i^j$ for fragment *i* which occur in the volume $V_i^j$ is simply proportional to the local partition function $Z_{i,j}$. Further, the number of configurations $k_i^j$ is precisely the number of neighbors of configuration *j* for fragment *i*. Therefore, the ratio of partition functions occurring in Eq. 28 is equal to the ratio of the numbers of configurations in the corresponding neighbor lists, and the acceptance criterion simplifies to

$$p_{acc}(o \to n) = \min\left[1, \exp(-\beta \Delta U^{\text{rest}}) \frac{k_i^o}{k_i^n}\right]. \tag{29}$$

Eq. 29 is indeed the same as Eq. 13 for the neighbor list swap moves derived using the re-weighing approach, completing our alternate derivation.

**Acknowledgements**

We would like to thank Marty Ytreberg, Ed Lyman, Ying Ding, Bin Zhang, Xin Zhang, Andrew Petersen, and Eran Eyal for helpful discussions. Funding was provided by the NIH through grants GM070987 and GM076569, as well as by the NSF through grant MCB-0643456.

**Tables**

| Protein | Number of residues | Time |
|---|---|---|
| Protein G | 56 | 20 min |
| Calmodulin | 71 | 8.64 min |
| Barstar | 89 | 47 min |
| CDC25B | 177 | 9 h |
| GGBP | 309 | 24 h |

Table 1. Decorrelation time for five different proteins in units of wallclock time of a single CPU using a method described in Ref. (24).

**Figure legends**

Figure 1. The protein backbone model used in this study is represented by a set of peptide plane configurations. A library of atomistic peptide-plane configurations is generated in advance and used for trial moves in library-based Monte Carlo, allowing full flexibility of the atomic backbone.

Figure 2. The one-dimensional toy system used as a "reality check" of library-based Monte Carlo (LBMC) is shown schematically in (A) and consists of two particles connected to harmonic springs and restrained in one-dimension. The equilibrium position of the first particle was set at the origin and the second 3 Å away from the origin. The probability density function (pdf) along the one-dimensional coordinate calculated LBMC and checked with BD for two cases: (B) particles do not interact with each other, and (C) particles interact with each other via Coulombic repulsion. BD results of the first particle are denoted by black line with circles and the second particle by red line with squares. LBMC results are denoted by green line with diamonds for the first particles and by blue line with triangles for the second particle. Note that the equilibrium distance between particles increased when the Coulombic repulsion was switched on (C) which was correctly reproduced by LBMC. Both particles were restrained with a spring constant of 0.5 kcal/(mol*Å$^2$) at the temperature of 300 K. In (C) the partial charges of particles were set to 0.2 e and the dielectric constant was set to 1.

Figure 3. Root mean square deviation (RMSD) along the library-based Monte Carlo (LBMC) trajectory relative to the experimental structure for five different proteins: (A) protein G, (B) calmodulin, (C) barstar, (D) CDC25B, and (E) GGBP. RMSD was calculated based on all backbone heavy atoms. For CDC25B (D) RMSD was calculated based on the whole protein (black) and only on the stable part of the protein (residues 374-524) (red). These RMSD plots show that the LBMC technique is capable of sampling large conformational fluctuations along with apparent convergence characterized by stable behavior. The simulations were performed in less than three weeks on a single commercial processor.

Figure 4. Convergence analysis of LBMC simulations for five different proteins: (A) protein G, (B) calmodulin, (C) barstar, (D) CDC25B, and (E) GGBP. Each plot shows the convergence properties analyzed using the procedure described in Ref. (24). *N* denotes the interval between frames along the trajectory at which the convergence properties are calculated. The number of frames required for the normalized variance ($\sigma^2$) to reach the value of one (horizontal line) is an approximation of the structural decorrelation time. The three curves on each plot are results for different subsample size and demonstrate the robustness of the value for the decorrelation time; see Ref. (24). The error bars correspond to estimates of an 80% confidence interval intrinsic to the number of subsamples studied for the green line.

Figure 5. Superposition of 20 different backbone configurations of CDC25B along the LBMC trajectory. The stable part of the protein (residues 374-524) is shown in cyan and the flexible C-terminus helix in red. Interestingly, isoform A of CDC25 has a similar structure except for the C-terminus helix that is unfolded (42). In CDC25B this helix forms a cleft running along the protein body with an anion binding site at the end of the helix occupied by Cl$^-$. Since the protein molecules in the crystal are in contact by C-termini these helices may be a crystal structure artifact caused by crystal packing forces and favorable electrostatic interactions. [Figure produced using the program VMD (48)].

**Figures**

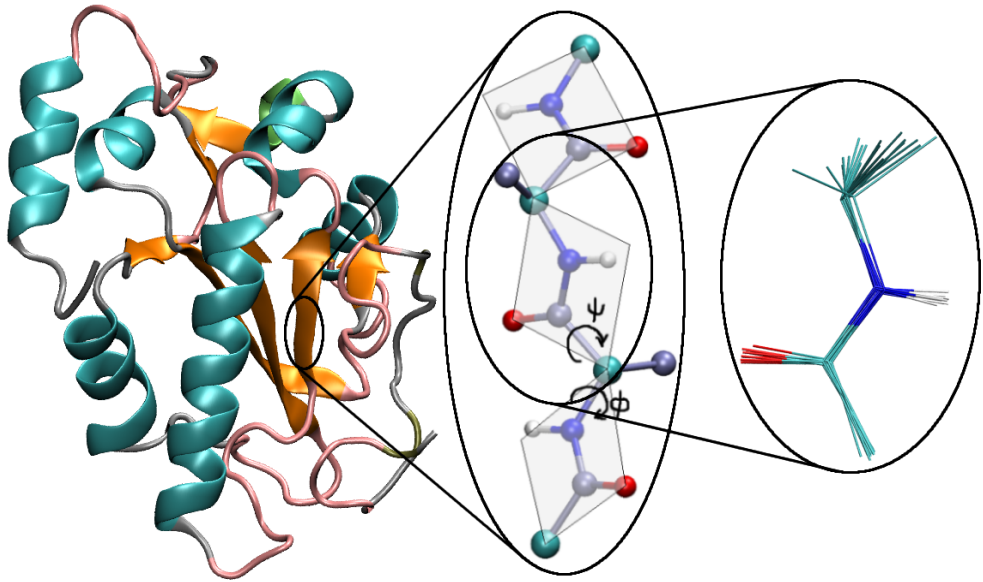

Figure 1.

A

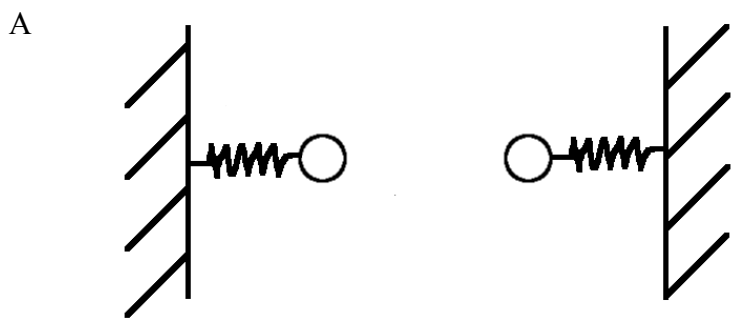

B

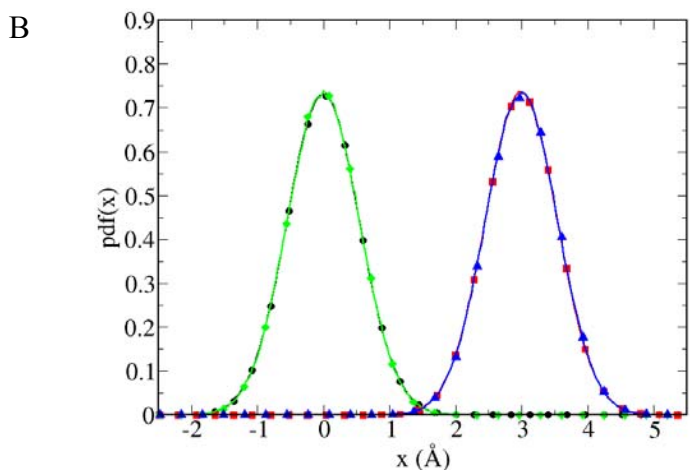

C

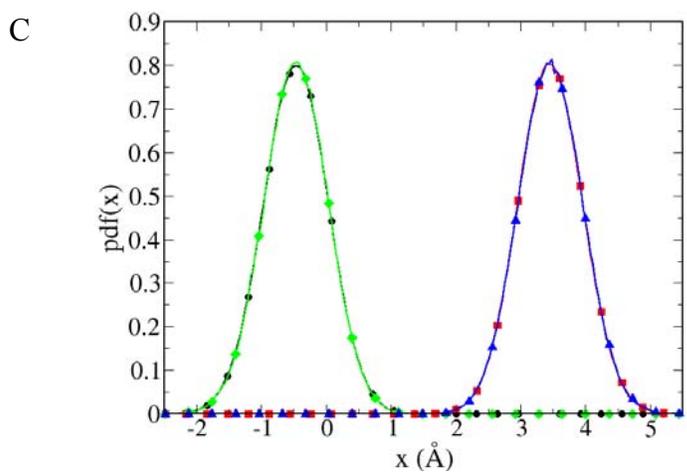

Figure 2.

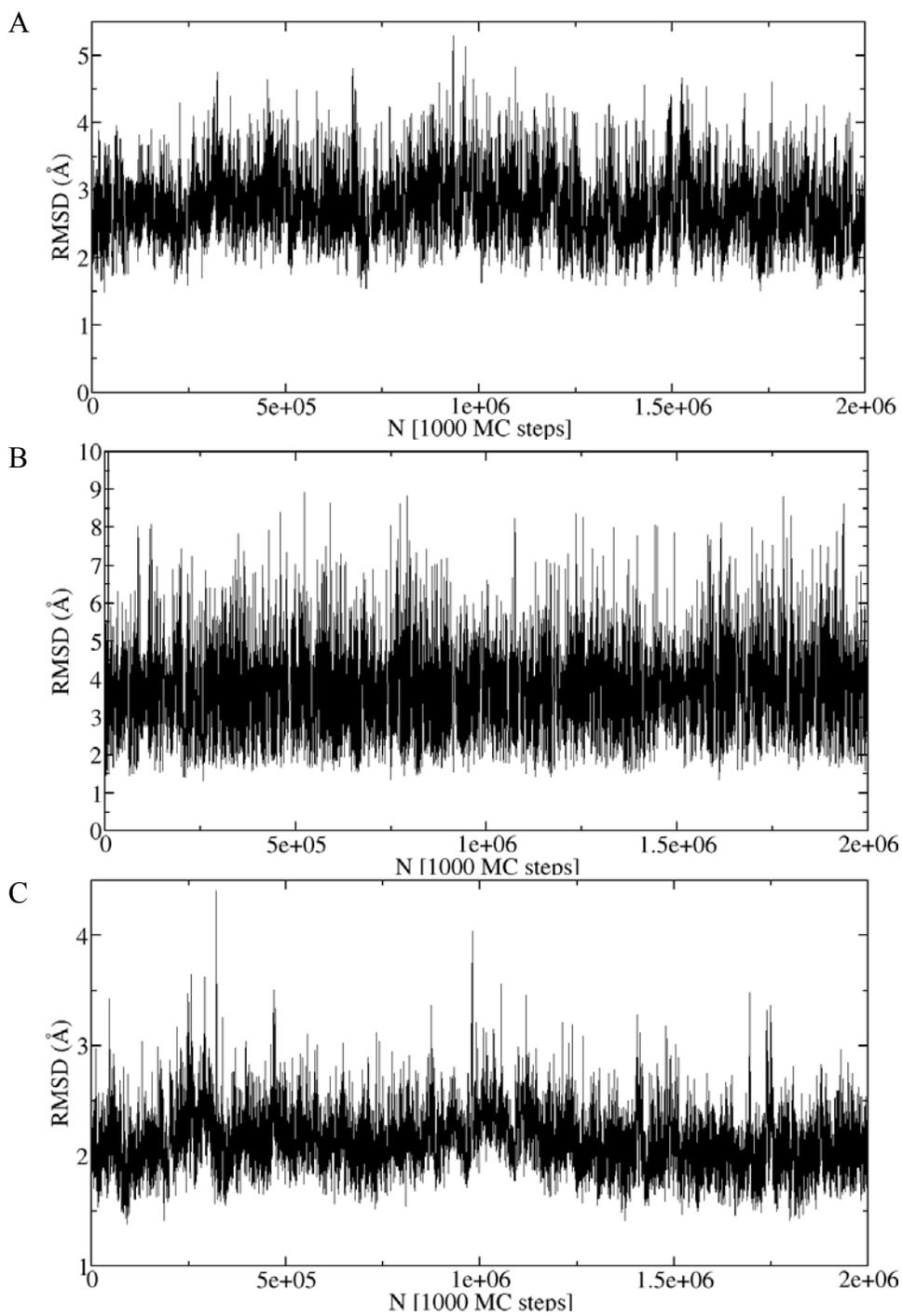

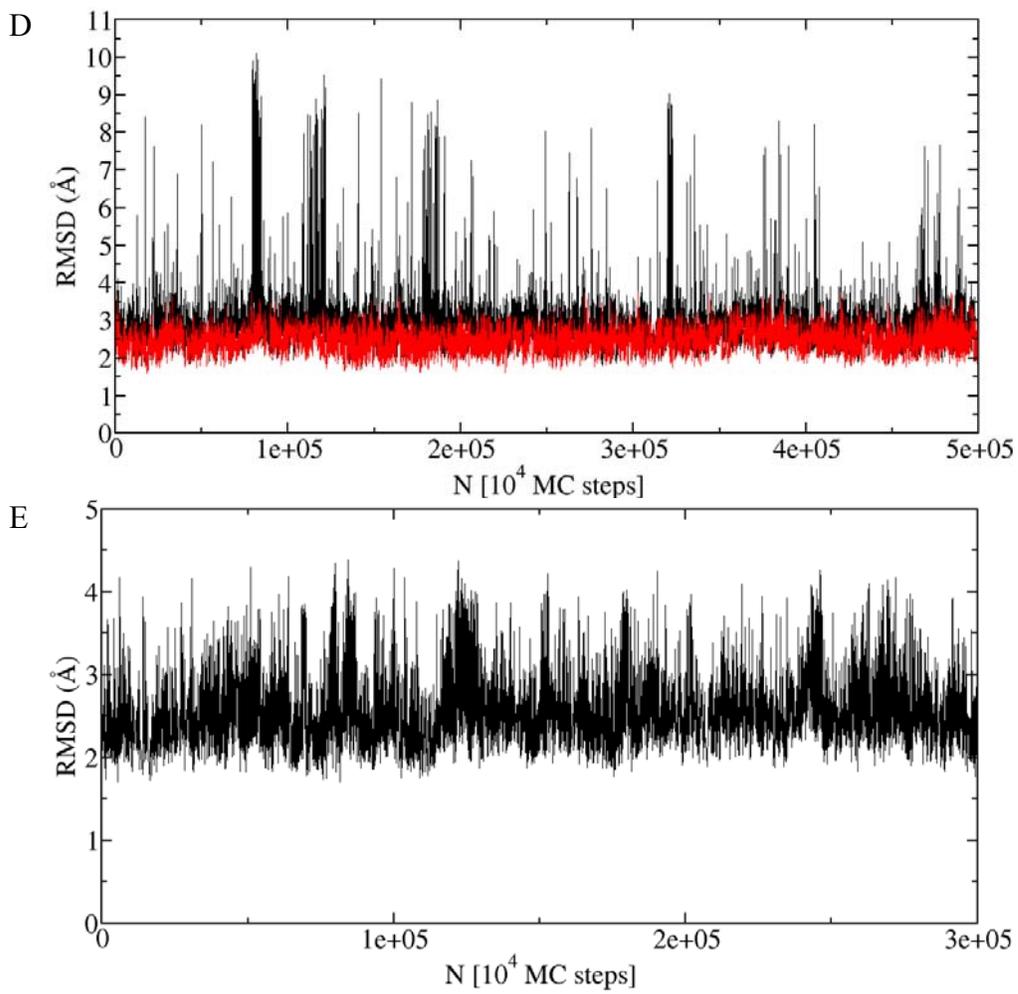

Figure 3.

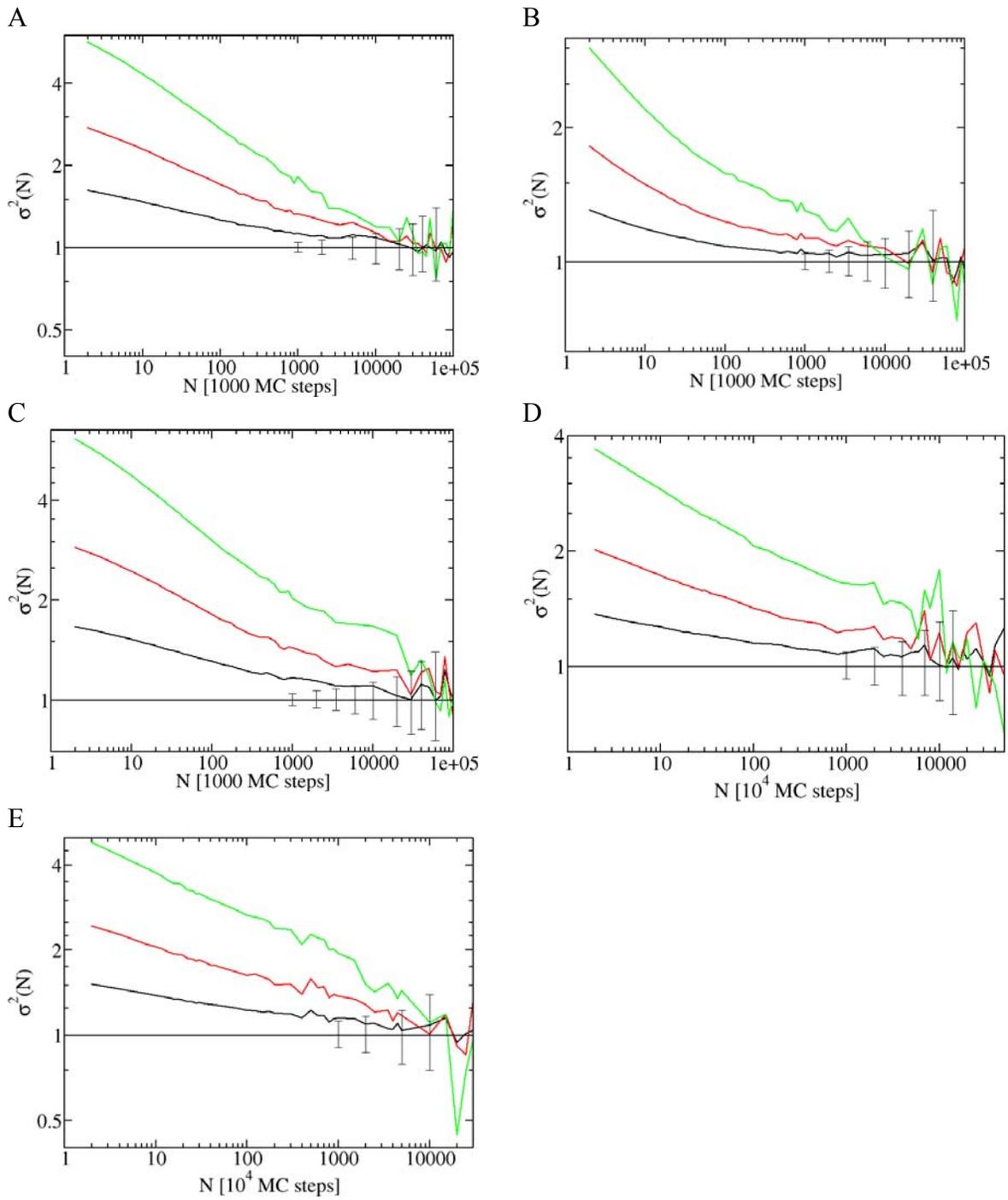

Figure 4.

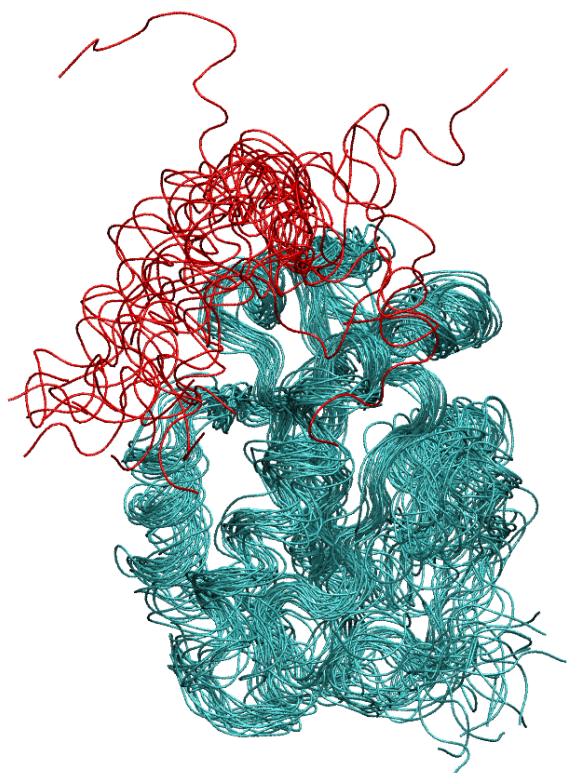

Figure 5.